\documentclass[10pt,journal,compsoc]{IEEEtran}

\ifCLASSOPTIONcompsoc
  \usepackage[nocompress]{cite}
\else
  \usepackage{cite}
\fi

\ifCLASSINFOpdf
\else
\fi

\usepackage{amsmath,amssymb}
\usepackage{changepage}
\usepackage{tikz}
\usetikzlibrary{positioning, calc}
\usetikzlibrary{shapes.geometric}
\usepackage[flushleft]{threeparttable}
\usepackage{array,booktabs,makecell}
\usepackage{setspace}
\usepackage{algorithm}%
\usepackage{algpseudocode}%
\DeclareMathOperator*{\argmin}{argmin}
\usepackage{graphicx}

\makeatletter
\newcommand{\subalign}[1]{%
  \vcenter{%
    \Let@ \restore@math@cr \default@tag
    \baselineskip\fontdimen10 \scriptfont\tw@
    \advance\baselineskip\fontdimen12 \scriptfont\tw@
    \lineskip\thr@@\fontdimen8 \scriptfont\thr@@
    \lineskiplimit\lineskip
    \ialign{\hfil$\m@th\scriptstyle##$&$\m@th\scriptstyle{}##$\hfil\crcr
      #1\crcr
    }%
  }%
}
\makeatother

\usepackage{physics}
\usepackage{cancel}
\newcolumntype{C}[1]{>{\centering\arraybackslash}p{#1}}
\newcolumntype{L}[1]{>{\raggedright\arraybackslash}p{#1}}

\usepackage{amsthm}
\usepackage{thmtools}
\declaretheoremstyle[
spaceabove=6pt, spacebelow=6pt,
headfont=\normalfont\bfseries,
notefont=\mdseries, notebraces={(}{)},
bodyfont=\normalfont,
postheadspace=0.6em,
headpunct=:
]{mystyle}

\usepackage{url}
\usepackage{tabularx}
\makeatletter
\newcommand{\multiline}[1]{%
  \begin{tabularx}{\dimexpr\linewidth-\ALG@thistlm}[t]{@{}X@{}}
    #1
  \end{tabularx}
}

\tikzset{
    buffer/.style={
        draw,
        shape border rotate=0,
        regular polygon,
        regular polygon sides=3,
        fill=blue,
        node distance=.05ex, %
        minimum height=.05ex %
    }
}

\usepackage{graphicx}
\usepackage{xcolor}

\newcommand{\kmax}{{k_{\textrm{max}}}}
\newcommand{\jdam}{{\textrm{JDAM}}}

\newcommand{\metric}{{\text{mutual information}}} %
\newcommand{\metrictwo}{{\text{mutual information}}}
\newcommand{\joint}{{\mathcal{V}}}
\newcommand{\jointedge}{{\mathcal{E}}}
\newcommand{\obj}{{Q}}
\newcommand{\Ione}{{I_{\alpha}(q; q')}}
\newcommand{\Itwo}{{I_{\alpha}(q, m; q', m')}}
\newcommand{\Imetric}{{I_{\alpha}(m|q; m'|q')}}
\newcommand{\aone}{{\gamma_{\textrm{deg}}}}
\newcommand{\atwo}{{\gamma_{\textrm{att}}}}
\newcommand{\homoprob}{{\rho_{\textrm{att}}}}
\usepackage{array}
\usepackage{multirow}
\usepackage{amsfonts}
\usepackage{enumerate}
\usepackage{amsthm}
\usepackage{svg}
\usepackage{bbm}
\makeatletter
\algnewcommand{\LineComment}[1]{\Statex \hskip\ALG@thistlm \(\triangleright\) #1}
\makeatother

\begin{document}
\newcommand{\equilateral}{
\begin{tikzpicture}[scale=0.8]
    \coordinate[]  (A) at (0,0);
    \coordinate[] (B) at (.4,0);
    \coordinate[] (C) at (.2,.3464);
    \fill [blue] (A) -- (B) -- (C) -- cycle;
  \end{tikzpicture}
}
\title{Mutual Information Measure for Glass Ceiling Effect in Preferential Attachment Models}

\author{Rui~Luo,
        Buddhika~Nettasinghe,
        and~Vikram~Krishnamurthy,~\IEEEmembership{Fellow,~IEEE}%
\IEEEcompsocitemizethanks{\IEEEcompsocthanksitem R. Luo is with the Sibley School of Mechanical and Aerospace Engineering, Cornell University, Ithaca, NY, 14850.\protect\\
E-mail: rl828@cornell.edu
\IEEEcompsocthanksitem B. Nettasinghe is with Tippie College of Business, University of Iowa, Iowa City, IA 52242.\protect\\
E-mail: buddhika-nettasinghe@uiowa.edu%
\IEEEcompsocthanksitem V. Krishnamurthy is with the School of Electrical and Computer Engineering, Cornell University, Ithaca, NY, 14850.\protect\\
E-mail: vikramk@cornell.edu%
\IEEEcompsocthanksitem This research was supported in part by the  U. S. Army Research Office under grant W911NF-21-1-0093, and the National Science Foundation under grant CCF-2112457.\protect\\}}

\IEEEtitleabstractindextext{
\begin{abstract}
We propose a new way to measure inequalities such as the glass ceiling effect in attributed networks. Existing measures typically rely solely on node degree distribution or degree assortativity, but our approach goes beyond these measures by using mutual information (based on Shannon and more generally, Rényi entropy) between the conditional probability distributions of node attributes given node degrees of adjacent nodes. 
We show that this mutual information measure aligns with both the analytical structural inequality model and historical publication data, making it a reliable approach to capture the complexities of attributed networks. Specifically, we demonstrate this through an analysis of citation networks.
Moreover, we propose a stochastic optimization algorithm using a parameterized conditional logit model for edge addition, which outperforms a baseline uniform distribution. By recommending links at random using this algorithm, we can mitigate the glass ceiling effect, which is a crucial tool in addressing structural inequalities in networks.
\end{abstract}

}

\maketitle

\IEEEdisplaynontitleabstractindextext

\IEEEpeerreviewmaketitle

\section{Introduction}
\label{sec:introduction}
The glass ceiling effect is a hidden barrier that keeps certain groups of people from attaining positions of influence. 
Social networks are an important area where the glass ceiling effect arises. For instance, faculty hiring networks have been used to examine inequality and hierarchical structure in university prestige and gender \cite{clauset2015systematic}. Coauthorship of articles enables us to examine how researchers collaborate and what factors  have the strongest impact on that collaboration \cite{newman2004coauthorship}. 
A study on citation networks has shown the structural inequality that restricts the research impact of scientists from certain groups \cite{nettasinghe2021emergence}. Female researchers tend to be underappreciated in that their publications are published in less esteemed journals\cite{ross2020leaky} and tend to receive fewer citations\cite{dion2018gendered}.

In this paper, we use the mutual information measure (based on Shannon entropy and more generally Rényi entropy) to quantify the glass ceiling effect within an attributed network. Specifically, we show that the mutual information between the conditional probabilities of two nodes connected by a randomly chosen link offers valuable insights into the reduction of uncertainty between the attributes of the nodes, given their degrees. Here the conditional probability refers to the probability of observing specific attribute values for a node given its degree. 
This measure aptly reflects the complex dependencies in an attributed network and allows us to identify the presence of structural inequality. It helps identify whether a node's demographic traits, such as gender, have an impact on the recognition or attention she receives within the network. 

Compared to other approaches for quantifying network inequality (reviewed below), the proposed mutual information measure offers the following advantages: 
\begin{itemize}
    \item It captures general dependencies across the entire degree distribution, as opposed to degree assortativity which only accounts for linear relationships. 
    \item Additionally, it enables a computationally efficient optimization algorithm by transforming the discrete optimization into a stochastic optimization over the probability space of edge addition.
\end{itemize}
As a result, the proposed method is useful both analytically and computationally.

\vspace{0.1in}
\noindent
{\bf Main Results and Organization:}

\noindent (1) Section \ref{sec:information} introduces the \metric{} as a measure that quantifies the information related to node attributes in an attributed network. In Section \ref{subsec: jdam}, we propose a computationally efficient algorithm that uses a joint degree and attribute matrix (\jdam{}) to compute \metric{}.

\noindent (2) In Section \ref{subsec: why Delta I}, we justify the usefulness of \metric{} in measuring network inequality. We show that \metric{} outperforms both homophily and degree assortativity in capturing the information related to node attributes of a network using sample graphs. Furthermore, we use simulations to illustrate that the mutual information with Rényi entropy of order $\alpha=1.3$ is the most expressive in terms of capturing the node attribute information.

\noindent (3) The mutual information measure (based on Shannon or more generally Rényi entropy) is not a submodular function with respect to adding edges, thus a discrete optimization for individual edges is intractable. In Section \ref{sec:optim}, we relax the problem into a stochastic optimization over the space of probability mass functions of adding edges. 
The resulting conditional logit model offers recommendations for links that will enhance network inequality.

\noindent (4) Section \ref{sec: numerical} uses simulations of a Directed Mixed Preferential Attachment (DMPA) model together with real-world citation networks from various fields to show how \metrictwo{} quantifies gender-based glass ceiling effect in the research community.

\vspace{0.1in}
\noindent
{\bf Related Works:}

\noindent (1) \textit{Discrete choice in networks. } 
A collection of studies \cite{overgoor2019choosing, overgoor2020scaling, gupta2022mixed} examine the network formation process through the lens of discrete choice theory. Analysis of network inequality can result from the factors that influence decision-making. 
Overgoor et al. \cite{overgoor2019choosing} described the network growth as a discrete choice for each node to form edges with others. They used the multinomial logit (MNL) model, which encompasses a variety of network features like preferential attachment and homophily, to describe the likelihood of selecting each alternative. Gupta and Porter \cite{gupta2022mixed} presented repeated-choice (RC) model to study network formation. The RC model relaxes certain limitations of the MNL model, such as the assumption that each node makes exactly one choice and shares the choice set. 
To account for effects the degree and attribute assortativity has on the network formation, Sadler \cite{sadler2020diffusion} devised a diffusion game in which each node choose whether to adopt a new behavior. To show how the new behavior diffused, they used a branching process on a random network with nodes drawn based on a configuration model. 

\vspace{0.1in}
\noindent (2) \textit{Glass ceiling effect in networks.} 
Various discrete-time network models \cite{avin2015homophily, nettasinghe2021emergence, nettasinghe2022scale} have been proposed to depict the dynamics of the glass ceiling effect. 
In our study, the proposed \metrictwo{} is computed based on each network snapshot, which  depicts the evolution of the glass ceiling effect over time and is well suited to existing models. 
Avin et al. \cite{avin2015homophily} proposed a preferential attachment model of undirected networks for the glass ceiling effect, where preferential attachment, homophily, and minorities are necessary and sufficient conditions for the emergence of glass ceiling effect. 
Nettasinghe et al. \cite{nettasinghe2021emergence, nettasinghe2022scale} devised the Directed Mixed Preferential Attachment (DMPA) model, which incorporates directed preferential attachment and homophily into the growth dynamics of a directed graph that contains a minority and a majority. They analyzed the in-degree and out-degree distributions of the minority and majority groups under the DMPA model. The DMPA model explores the glass ceiling effect via the interplay between the structure and dynamics of directed networks.

\section{Mutual Information of attributed networks}\label{sec:information}
This section begins by introducing the key concepts of attributed networks, remaining degree distribution, and the joint remaining degree distribution of adjacent nodes, which are essential for defining the information content of a network. We then propose the \metric{}, that incorporates node attributes to measure the information content of an attributed network. We then introduce an extension of \metric{}, \metrictwo{}, and demonstrate its superior correlation coefficient with node attribute assortativity. 
This analysis offers critical insights into the role of node attributes in network edge formation and emphasizes the importance of incorporating these attributes into network inequality analysis. Our proposed mutual information measure provides a valuable tool for assessing the impact of node attributes on network structure and can aid in mitigating structural inequalities in networks.

\begin{figure}
	\centering
	\includegraphics[width=0.4\textwidth]{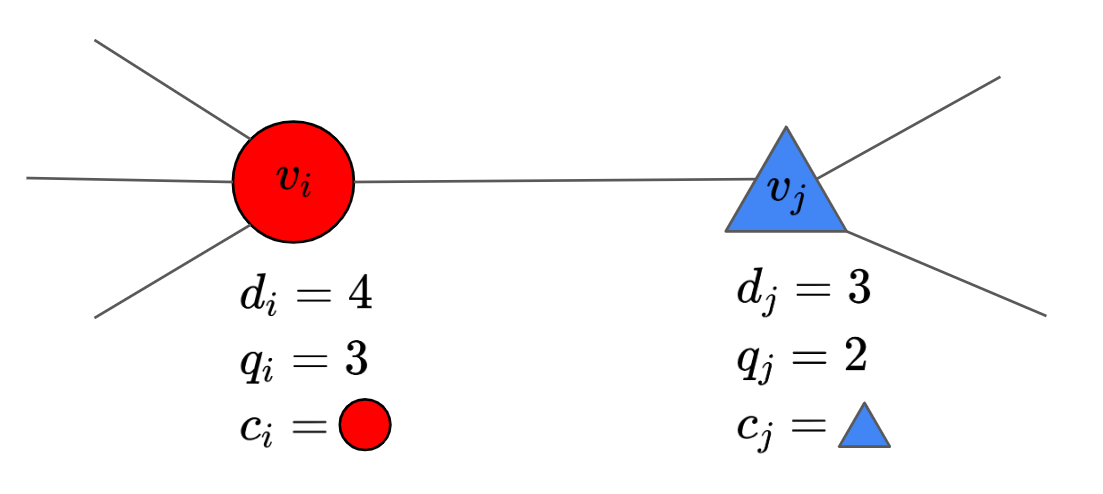}
	\caption{We display the degrees $d$, the remaining degrees $q$, and the node attributes $c$ of two connected nodes $v_i$ and $v_j$ in an attributed network. The remaining degree and the joint remaining degree follow the distribution $q_k$ (\ref{eq:remaining degree}) and $e_{kk'}$, respectively. The degree assortativity of the network is defined in (\ref{eq: aone}), whereas the network's mutual information based on the remaining degree distribution is defined in (\ref{eq: degree MI}).}
	\label{fig: degree correlation}
\end{figure}

\subsection{Definition of an Attributed Network}
\label{subsec: definition of coauthorship network}
We define an attributed network as a labeled undirected multigraph $G =(V, E, w, C)$, where $V=\{v_i\}_{i=1, \cdots, n}$ is the set of $n$ authors; $E=\{\{ v_i, v_j \}\}$ is the set of undirected edges\footnote{We ignore the edge direction for expositional simplicity in defining the $\jdam$ (Section \ref{subsec: jdam}) and assortativity (Section \ref{subsec: why Delta I}).} between nodes; $w: E \rightarrow \mathbb{Z}^+$ denote the edge multiplicities, each representing the number of times that the two nodes have interacted with each other; $C=\{c_i\}_{i=1, \cdots, n}$ assign a binary label $c_i \in \{+1, -1\}, i=1,\cdots, N$ to each node based on a demographic attribute such as gender\footnote{We can do the partition based on other attributes as well in citation networks, such as the prestige of authors' institutional affiliation\cite{nettasinghe2021emergence} and the majority/minority identity.}. 
The degree of a node $v_i$ is the sum of the multiplicities of the edges attached to it: %
$d_i = \sum_{\{v_i, v_j\} \in E} w(\{v_i, v_j\})$. 
The degree distribution is defined as $(p_k)_{k=1, \cdots, \kmax}$, where $p_k$ denotes the probability that a randomly chosen node in the graph will have degree $k$. We are interested here in the remaining degree distribution\cite{newman2002assortative} defined as 
\begin{equation}\label{eq:remaining degree}
    q_k = \frac{(k+1)p_{k+1}}{\sum_{j=1}^{\kmax} j p_j},  k \in \{0, \cdots, \kmax - 1 \} .
\end{equation}
which quantifies the distribution of the number of edges leaving one node other than the edge that we arrived along.

We now define $e_{kk'}$ to be the joint probability distribution of the remaining degrees of the two nodes at either end of a randomly chosen edge. This quantity is symmetric in that $e_{kk'} = e_{k'k}$ and it obeys the sum rules $\sum_{kk'} e_{kk'} = 1$ and $\sum_{k} e_{kk'} = q_{k'}$\cite{newman2002assortative}. 

We also define the attribute distribution as $m(c), c\in \{+1, -1\}$. The joint distribution of the attributes of two nodes incident to an edge is $m(c, c'), c, c'\in \{+1, -1 \} $.  Figure \ref{fig: degree correlation} shows the statistics of a pair of nodes connected by an edge.

\subsection{Information Content of an Attributed Network}
\label{subsec: jdam}
In this subsection, we define the information content of attributed networks, where network degree information is publicly available, such as the number of citations an author has. However, node attributes, such as gender, research area, or academic rank, may not be entirely determined by node degrees. To address this, we introduce the \metric{} framework, which quantifies the reduction in uncertainty in node attributes relative to node degrees. 

\subsubsection{Shannon Mutual Information}
The Shannon entropy\footnote{We assume logarithm to base 2 throughout the paper.} of a network with respect to (\ref{eq:remaining degree}) is defined as:
\begin{equation} \label{eq:Shannon entropy}
    H(q) = -\sum_{k=0}^{\kmax-1} q_k \log(q_k)
\end{equation}
which quantifies the heterogeneity of the network's degree distribution. 

Furthermore, the network's information content \cite{piraveenan2009assortativeness} (also referred to as information transfer in \cite{sole2004information}) can be determined by considering the remaining degree distribution $q'$ of the node at the other end of a randomly chosen link. Specifically, the information content is defined as:
\begin{equation} \label{eq: basic MI}
I(q;q') = H(q) - H(q|q'),
\end{equation}
where %
\begin{equation}
\label{eq: degree MI}
    \begin{split}
        H(q|q') & = \sum_{k'=0}^{\kmax-1} q_{k'} \sum_{k=0}^{\kmax-1} \pi_{k|k'} \log \frac{1}{\pi_{k|k'}} \\
        & = \sum_{k'=0}^{\kmax-1} \sum_{k=0}^{\kmax-1} e_{kk'} \log \frac{q_{k'}}{e_{kk'}}
    \end{split}
\end{equation} 
is the conditional entropy defined via conditional probabilities $\pi_{k|k'} = \frac{e_{kk'}}{q_{k'}}$ of observing a node with $k$ edges leaving it provided that the node at the other end of the chosen edge has $k'$ leaving edges. In information-theoretic terms, $H(q|q')$ is equivocation (non-assortativity) between $q$ and $q'$, i.e., the assortative noise within the network's information channel\cite{prokopenko2009information}.

\begin{figure}
	\centering
	\includegraphics[width=0.5\textwidth]{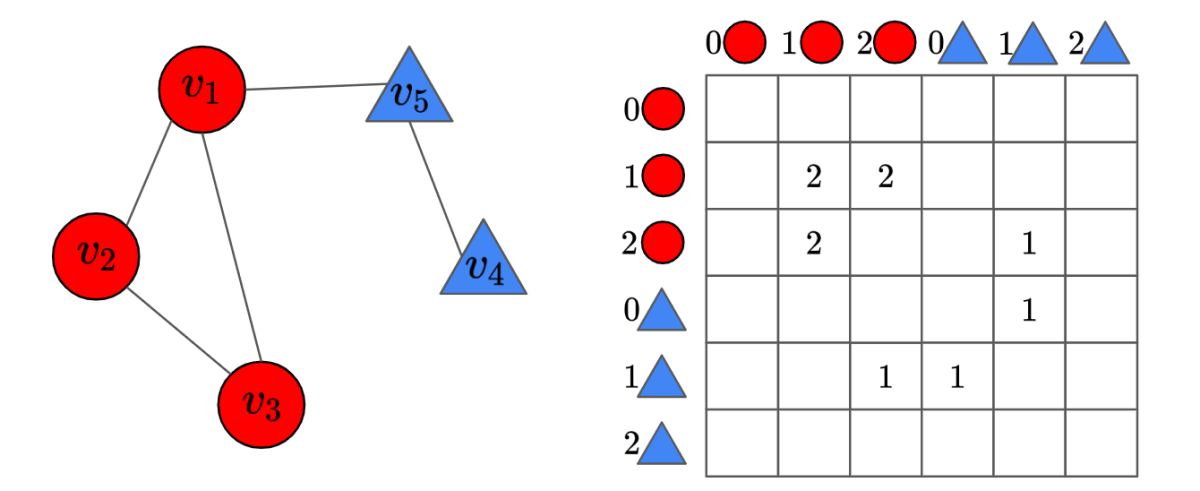}
	\caption{On the left, we show a citation network where the red circles represent male authors and blue triangles represent female authors, i.e., $c_1=c_2=c_3=m, c_4=c_5=f$. %
     On the right, the corresponding joint degree and attribute matrix (JDAM) (\ref{eq: basic JDAM}) is displayed, which is used to compute the information content (\ref{eq: basic MI}) and the \metric{} (\ref{eq: Delta I JDAM}).}
	\label{fig: joint degree and attribute}
\end{figure}

Given the entropy and the conditional entropy, the information content (\ref{eq: basic MI}) can be expressed as
\begin{equation}
\label{eq: degree MI expanded}
    I(q; q') = \sum_{k=0}^{\kmax-1} \sum_{k'=0}^{\kmax-1} e_{kk'} \log \frac{e_{kk'}}{q_k q_{k'}}
\end{equation}
which measures the amount of network degree correlations, i.e., the information of an author's degree conveyed by degrees of authors she cites.

\subsubsection{Rényi Mutual Information}
The Rényi entropy \cite{fehr2014conditional, verdu2015alpha} is a generalization of Shannon entropy and has many applications in fields such as unsupervised learning, source adaptation, and image registration. The Rényi entropy of order $\alpha$ is defined as follows for the node remaining degree distribution $q$: 
\begin{equation}
    H_{\alpha}(q) = \frac{1}{1-\alpha} \log(\sum_{k=0}^{\kmax-1} q_k^{\alpha}), \nonumber
\end{equation}
where $\alpha \geq 0$ and $\alpha \neq 1$.
The limiting value of ${H} _{\alpha }$ as $\alpha \rightarrow 1$ is the Shannon entropy (\ref{eq:Shannon entropy}). The conditional Rényi entropy (\ref{eq: degree MI}) is defined following $H_{\alpha}(q|q') = H_{\alpha}(q, q') - H_{\alpha}(q')$:
\begin{equation}
    H_{\alpha}(q|q') = \frac{1}{1-\alpha} \log \left( \frac{\sum_{k'=0}^{\kmax -1} \sum_{k=0}^{\kmax-1} e_{kk'}^{\alpha}}{\sum_{k'=0}^{\kmax-1} q_{k'}^{\alpha}} \right) \nonumber
\end{equation}

Replacing the Shannon entropy in the definition of \metric{} (\ref{eq: basic MI}) by the Rényi entropy, one obtains the Rényi mutual information:
\begin{equation}
\begin{split}
    I_{\alpha}(q;q') &= H_{\alpha}(q) - H_{\alpha}(q|q') \\ &= \frac{1}{1-\alpha} \log \left( \frac{\sum_{k=0}^{\kmax-1} q_{k}^{\alpha} \sum_{k'=0}^{\kmax-1} q_{k'}^{\alpha}}{\sum_{k=0}^{\kmax -1} \sum_{k'=0}^{\kmax-1} e_{kk'}^{\alpha}} \right)
    \nonumber
\end{split}
\end{equation}

Now taking the node attributes into account, we define the joint mutual information as 
\begin{equation}
    \Itwo = H_{\alpha}(q, m) - H_{\alpha}(q, m|q', m'), \nonumber
\end{equation} 
which is the mutual information between the two bivariate variables $(q, m)$ and $(q', m') $
that stand for the degrees and the attributes of a pair of two adjacent nodes. 
It can be expanded as 
\begin{equation}
\begin{split}
    & \Itwo \\ = &\frac{1}{1-\alpha} \log \left( \frac{\sum\limits_{k=0}^{\kmax-1} \sum\limits_{k'=0}^{\kmax-1} \sum\limits_{\substack{c\in \{+1, -1\} \\ c'\in \{+1, -1\}}} p(k, c)^{\alpha} p(k', c')^{\alpha}}{\sum\limits_{k=0}^{\kmax-1} \sum\limits_{k'=0}^{\kmax-1} \sum\limits_{\substack{c\in \{+1, -1\} \\ c'\in \{+1, -1\}}} p(k, c, k', c')^{\alpha}} \right) \nonumber    
\end{split}
\end{equation}

The proposed \metric{} measure is defined as the conditional mutual information of the attributes of neighboring nodes given the degrees of the nodes
\begin{equation}
\label{eq: Delta I basic}
\begin{split}
    I_{\alpha} & = \Imetric \\
    & = \Itwo - \Ione,
\end{split}
\end{equation}
which the mutual information is specifically defined between the conditional probability distributions of node attributes given node degrees of adjacent nodes.

We can rewrite (\ref{eq: Delta I basic}) as
\begin{equation}
\label{eq: Delta I advanced}
\begin{split}
    &I_{\alpha} \\ =& \frac{1}{1-\alpha} \log \left( \frac{\sum\limits_{k=0}^{\kmax-1} \sum\limits_{k'=0}^{\kmax-1} \sum\limits_{\substack{c\in \{+1, -1\} \\ c'\in \{+1, -1\}}} e_{kk'}^{\alpha} p(k, c)^{\alpha} p(k', c')^{\alpha}} {\sum\limits_{k=0}^{\kmax-1} \sum\limits_{k'=0}^{\kmax-1} \sum\limits_{\substack{c\in \{+1, -1\} \\ c'\in \{+1, -1\}}} q_k^{\alpha} q_{k'}^{\alpha} p(k, c, k', c')^{\alpha}} \right) 
\end{split}
\end{equation}

The Shannon mutual information is the special case of (\ref{eq: Delta I advanced}) when $\alpha \rightarrow 1$:
\begin{equation}
\label{eq: Shannon Delta I}
\begin{split}
    I &= \sum_{k=0}^{\kmax-1} \sum_{k'=0}^{\kmax-1} \sum_{\substack{c\in \{+1, -1\} \\ c'\in \{+1, -1\}}}  \Bigg( p(k, k', c, c') \\
    & \phantom{\sum_{k=0}^{\kmax-1} \sum_{k'=0}^{\kmax-1} \sum_{\scriptscriptstyle c\in \{+1, -1\}} \sum_{\scriptscriptstyle c'\in \{+1, -1\}}} \log \frac{q_k q_{k'} p(k, k', c, c')}{e_{kk'} p(k, c) p(k', c')} \Bigg) 
\end{split}
\end{equation}

$I_{\alpha}$ quantifies the increase in information content of an attributed network due to the inclusion of node attributes. 
In other words, $I_{\alpha}$ provides insight into the additional information that node attributes can offer beyond what can be explained by degree assortativity alone, thereby enabling a more accurate assessment of the underlying network inequality and glass ceiling effect.

To visualize $I_{\alpha}$ for a network, we first define the degree-attribute group $B_{kc} = \{v_i | q_i = k, c_i = c \} $ as the set of nodes that have degree $k$ and attribute $c$. There are at most $2\kmax $ degree-attribute groups, corresponding to each possible combination of degree and attribute values. 
Following \cite{gjoka2015construction}, we define the joint degree and attribute matrix ($\jdam$) as the number of edges connecting nodes in different degree-attribute groups $(k, c)$ and $(k', c')$:
\begin{equation}
\label{eq: basic JDAM}
    \jdam((k, c), (k', c')) = \sum_{\subalign{i: q_i &=k,\\ c_i &=c}} \sum_{\subalign{j: q_j &=k',\\ c_j &=c'}}  w(\{v_i, v_j\}),
\end{equation}
which capture the co-occurrence of the degrees and node attributes, and the entry in the normalized (normalized to sum 1) $\hat{\jdam}$ is $p(k, k', c, c')$. Figure \ref{fig: joint degree and attribute} shows an example citation network and its corresponding $\jdam$.

If we group the rows of different node attributes by node degree, we can get the joint degree distribution:
\begin{equation}
\begin{split}
    e_{kk'} & \propto \jdam(k, k') \\
     & = \sum_{i: q_i =k} \sum_{\subalign{j: q_j &=k'}}  w(\{v_i, v_j\}) \\
    & = \sum_{\substack{c\in \{+1, -1\} \\ c'\in \{+1, -1\}}} \jdam((k, c), (k', c')) \nonumber
\end{split}
\end{equation}
The normalized version of $\jdam$ can be represented as $\hat{\jdam}(k, k') = e_{kk'}$. Using $\hat{\jdam}$, $I_{\alpha}$ (\ref{eq: Delta I advanced}) can be computed as follows:
\begin{equation}
\label{eq: Delta I JDAM}
\begin{split}
    I_{\alpha} =& \frac{1}{1-\alpha} \log \Bigg( \sum_{k=0}^{\kmax-1} \sum_{k'=0}^{\kmax-1} 
    \sum_{\substack{c\in \{+1, -1\} \\ c'\in \{+1, -1\}}} 
    \hat{\jdam}(k, k')^{\alpha}  \\
    & \phantom{-----------} \hat{\jdam}(k, c)^{\alpha} \hat{\jdam}(k', c')^{\alpha} \Bigg) \\
    & -\frac{1}{1-\alpha} \log \Bigg( \sum_{k=0}^{\kmax-1} \sum_{k'=0}^{\kmax-1} 
    \sum_{\substack{c\in \{+1, -1\} \\ c'\in \{+1, -1\}}} \hat{\jdam}(k, k', c, c')^{\alpha} \\
   &\phantom{-----------} \hat{\jdam}(k)^{\alpha} \hat{\jdam}(k')^{\alpha} \Bigg) \nonumber
\end{split}
\end{equation}
Similarly, the \jdam{} representation of the Shannon mutual information (\ref{eq: Shannon Delta I}) is: 
\begin{equation}
\begin{split}
    I & =  \sum_{k=0}^{\kmax-1} \sum_{k'=0}^{\kmax-1} 
    \sum_{\substack{c\in \{+1, -1\} \\ c'\in \{+1, -1\}}} 
    \Bigg( \hat{\jdam}(k, k', c, c') \\
   &\phantom{=} \log \frac{\hat{\jdam}(k, k', c, c') \hat{\jdam}(k) \hat{\jdam}(k')}{\hat{\jdam}(k, c) \hat{\jdam}(k', c') \hat{\jdam}(k, k')} \Bigg) \nonumber
\end{split}
\end{equation}

\begin{figure}
	\centering
	\includegraphics[width=0.4\textwidth]{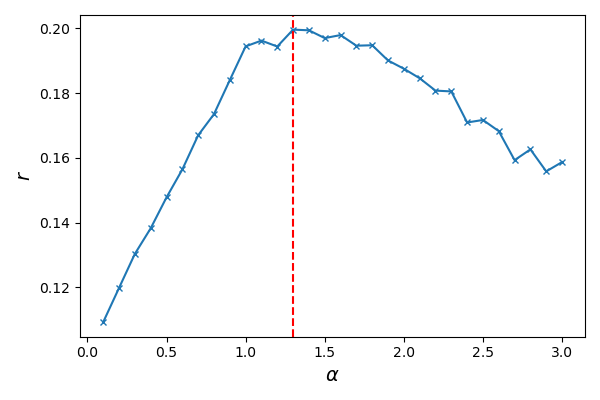}
	\caption{Simulations on stochastic block models (Section \ref{subsubsec: optimal order}) suggest that the Rényi mutual information of order $\alpha=1.3$ exhibits the highest correlation coefficient $r$ with $|\atwo|$. This finding provides insight on the optimal order $\alpha$ of Rényi mutualinformation that effectively captures node attribute information. Such an optimized choice of $\alpha$ will be incorporated into our optimization approach in Section \ref{sec:optim}.
 } 
	\label{fig: best alpha}
\end{figure}

\subsubsection{Optimal Order of Rényi Mutual Information}\label{subsubsec: optimal order}
In order to identify the optimal order $\alpha$ of Rényi mutual information for capturing node attribute information, we conducted a series of numerical experiments. To accomplish this, we fixed $n_1$ and $n_2$ nodes with attributes $+1$ and $-1$, respectively, and generated a network using a 2-block stochastic blockmodel (SBM) with a higher probability of intra-community connections than inter-community connections. We conducted multiple simulations, and for each simulated network, we permuted the node numbering and recalculated $\Ione$, $\aone$, and $\atwo$ accordingly. Additionally, to ensure the simulated network was connected, we removed disconnected networks during the simulation process.

We then computed the correlation coefficient $r$ between $I_{\alpha}$ and the absolute value of the attribute assortativity $|\atwo|$. Our results indicate that the Rényi mutual information of order $\alpha=1.3$ has the highest correlation coefficient with $|\atwo|$, suggesting it is the most effective method for capturing node attribute information. These findings are presented in Figure \ref{fig: best alpha}, which illustrates the relationship between $\alpha$ and $r$.

Choosing an order $\alpha > 1$ which is close to Shanon's entropy captures information on all the elements on the distributions, while still emphasizing the elements with high probabilities, thereby accentuating the degree-attribute groups that have strong associations. Additionally, using an order greater than one enables the utilization of near-linear estimation algorithms, which can expedite the computation of mutual information for large-scale networks.

\subsection{Why Mutual Information to Identify Glass Ceiling Effect?}
\label{subsec: why Delta I}
The aim of this section is to justify why \metric{} (Shannon and Rényi) is a useful measure for inequalities in a network. We argue that, despite being often used as metrics to examine the glass ceiling effect in networks, assortativity and homophily do not sufficiently capture the information associated with an attributed network. The three claims that follow serve to highlight our conclusion:

\vspace{0.1in}
\noindent (1) Both degree and attribute assortativity cannot explain the information contained within a network on their own.

\noindent (2) $\Ione$ tends to increase with higher absolute value of the degree assortativity $|\aone|$ but does not capture the attribute assortativity $\atwo$. 

\noindent (3) While $\Itwo$ is correlated with both degree and attribute assortativity, $I_{\alpha}$ reduces the effect that the network structure has on the network information and increases the information conveyed by the attributes. %

\begin{figure}
	\centering
	\includegraphics[width=0.45\textwidth]{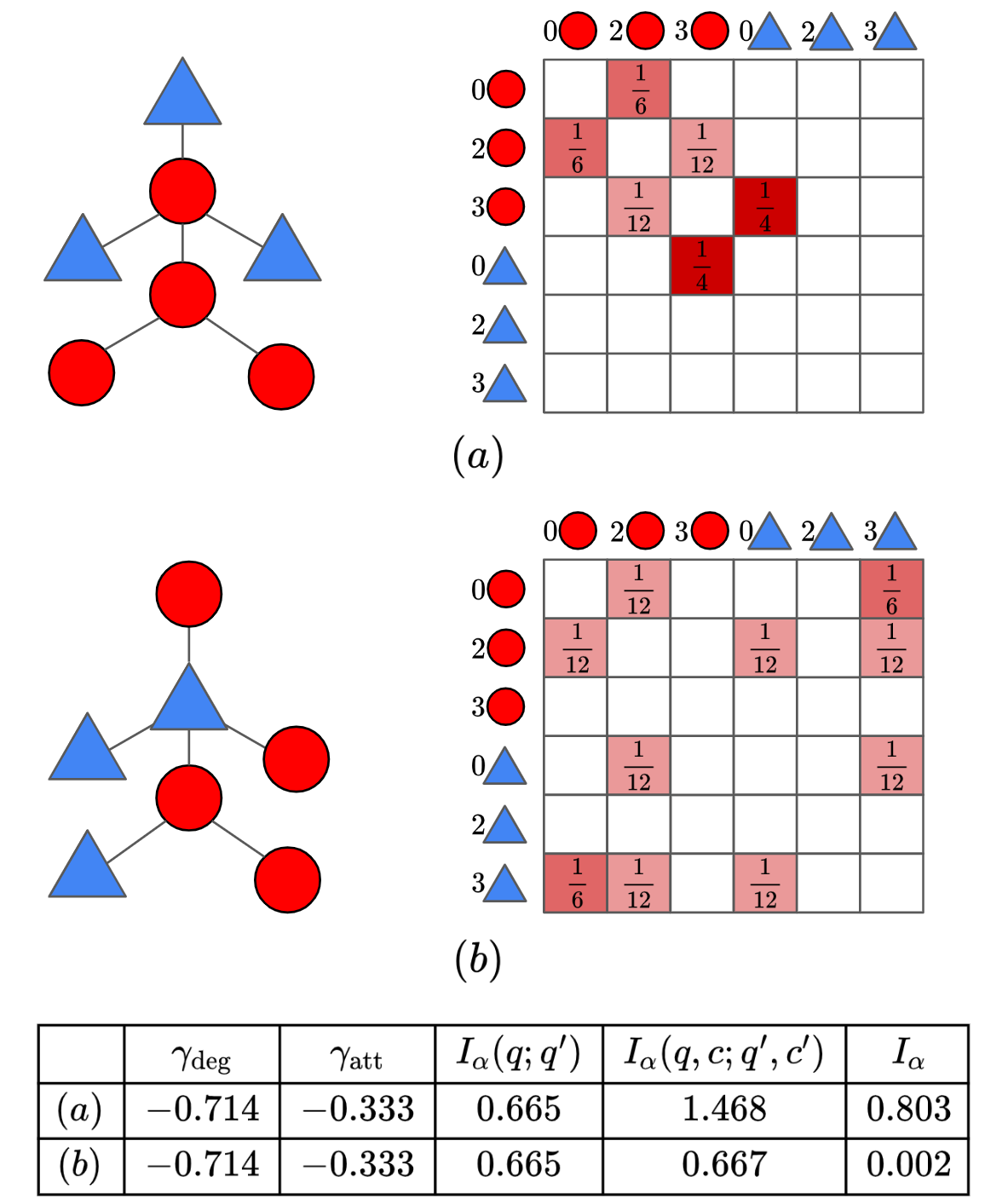}
	\caption{(Claim 1 of Section \ref{subsec: why Delta I}) Two networks $(a)$ and $(b)$ with identical topology but different node attributes with $\hat{\jdam}$ shown on the right. The bottom table includes the degree assortativity $\aone$, attribute assortativity $\atwo$, mutual information $\Ione$, joint mutual information $\Itwo$, and the proposed $I_{\alpha}$. By only connecting to nodes with different attribute and degree in $(a)$, the female nodes (represented by blue triangles) demonstrate perfect disassortativity, identified by the dark red cell in the corresponding $\hat{\jdam}$. In $(b)$, however, the female nodes are imperfectly disassortative which diffuses the $\hat{\jdam}$ and reduces $I_{\alpha}$ from 0.803 to 0.002. Degree assortativity and attribute assortativity may fail to convey the information associated with a network, as demonstrated in this example, where the proposed mutual information metric provides a more informative approach.}
	\label{fig: counter example}
\end{figure}

\vspace{0.2in}
\noindent
{\bf (1) Assortativity does not suffice.} 

Assortativity \cite{newman2003mixing} is a measure of how homophilic a graph is, and higher assortativity indicates stronger network inequality. The definition is slightly different according to whether the property of interest is categorical (attribute) or numeric (degree). The degree assortativity for undirected networks is defined as:
\begin{equation}
\label{eq: aone}
    \aone = \frac{\sum_{kk'} kk'(e_{kk'} - q_{k} q_{k'})}{\sigma_{q}^2}
\end{equation}
where $q_{k}$ and $e_{kk'}$ are the remaining degree distribution and the joint remaining degree distribution defined in Section \ref{subsec: definition of coauthorship network}, $\sigma_{q}$ is the standard deviation of the distribution $q_k$. $\aone$ measures the extent to which nodes with similar degree connect to each other (Eq. (24) in \cite{newman2003mixing}). 

The attribute assortativity is defined as: 
\begin{equation}
\label{eq: atwo}
    \atwo = \frac{\sum_{\scriptscriptstyle c\in \{+1, -1\}} p(c, c) - \sum_{\scriptscriptstyle c\in \{+1, -1\}} p(c) p(c)}{1 - \sum_{\scriptscriptstyle c\in \{+1, -1\}} p(c) p(c) }
\end{equation}
where $p(c)$ and $p(c, c')$ are the label distribution and the joint label distribution defined in Section \ref{subsec: definition of coauthorship network}. $\atwo$ measures the similarity of connections in the graph with respect to the node attribute (Eq. (2) in \cite{newman2003mixing}).

First of all, the degree assortativity (\ref{eq: aone}) does not consider the node attributes. Networks with the same topology but varied node attributes will have the same degree assortativity. Second, the attribute assortativity (\ref{eq: atwo}) calculates the \textit{global} linear correlation coefficient of neighboring node pairs without taking the node degree into account, which could have a \textit{local} impact on the connection pattern. In this context, \textit{local} refers to a particular degree-attribute region in the $\jdam$.

\begin{figure}
	\centering
	\includegraphics[width=0.5\textwidth]{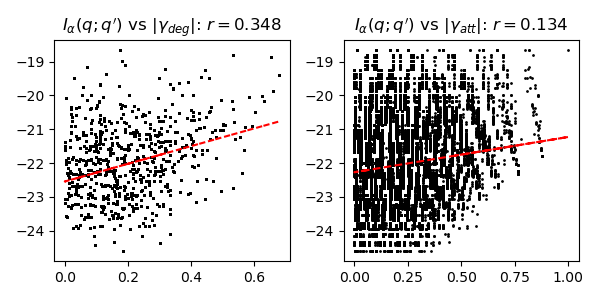}
	\caption{(Claim 2 of Section \ref{subsec: why Delta I}) The scatter plot illustrates a positive correlation between $\Ione$ and both $|\aone|$ and $|\atwo|$, indicating that greater levels of assortativity or disassortativity leads to higher $\Ione$ values. This indicates that higher levels of assortativity or disassortativity are associated with higher $\Ione$ values.} 
	\label{fig: I1 correlation}
\end{figure}

Figure \ref{fig: counter example} displays two networks and their corresponding $\hat{\jdam}$ (\ref{eq: basic JDAM}). The color and the shape of a node is defined according to Figure \ref{fig: joint degree and attribute}. Since the two networks have the same topology, their degree assortativity is also the same. More intriguingly, they share the same attribute assortativity. However, $(a)$ has a significantly greater \metric{} than $(b)$. It is a consequence of the female nodes' complete disassortativity in $(a)$, which causes a concentration in the $\hat{\jdam}$, as shown in the dark red cells of $(\{0, \equilateral{}\}, \{2, \tikz\draw[red,fill=red] (0,0) circle (0.8ex);\})$ and $(\{2, \tikz\draw[red,fill=red] (0,0) circle (0.8ex);\}, \{0, \equilateral{}\})$. On the other hand, the incomplete disassortativity in $(b)$ causes a leakage of the \metric{}, i.e., values in the $\hat{\jdam}$ diffuse to more cells.

\vspace{0.2in}
\noindent
{\bf (2) } $\mathbf{\Ione}$ {\bf increases with } $\boldsymbol{|\aone|}$ {\bf but does not reflect } $\boldsymbol{\atwo}.$

To demonstrate the correlation between $\Ione$ and both degree assortativity $\aone$ and attribute assortativity $\atwo$, we simulated networks from SBM according to Section \ref{subsubsec: optimal order}. Our observations reveal that networks with high $\Ione$ tend to possess a broader degree distribution and degree disassortativity.

In Figure \ref{fig: I1 correlation}, we illustrate the correlation between $\Ione$ and $\aone$ and $\atwo$. $\Ione$ is positively correlated with the absolute value of the degree assortativity $|\aone|$, i.e., $\Ione$ tends to increase as the level of degree assortativity or disassortativity in networks increases. This outcome aligns with previous observations on real-world networks presented in \cite{sole2004information} and the analytical model for scale-free networks presented in \cite{piraveenan2009assortativeness}. On the other hand, our results demonstrate that $\Ione$ has almost no correlation with $\atwo$, suggesting that node attribute information is not incorporated in the network's information content $\Ione$. 

\vspace{0.2in}
\noindent
{\bf (3) } $\mathbf{\Itwo}$ {\bf increases with both assortativities and} $\boldsymbol{I}$ {\bf strengthens } $\boldsymbol{\atwo}.$

Similar to $\Ione$, $\Itwo$ is positively correlated with $|\aone|$. Furthermore, $\Itwo$ has a positive correlation with $|\atwo|$, indicating that $\Itwo$ captures the information associated with the node attributes, and it tends to rise for both attribute-assortative (i.e., homophilic) and attribute-disassortative (i.e., heterophilic) networks.

\begin{figure}
	\centering
	\includegraphics[width=0.5\textwidth]{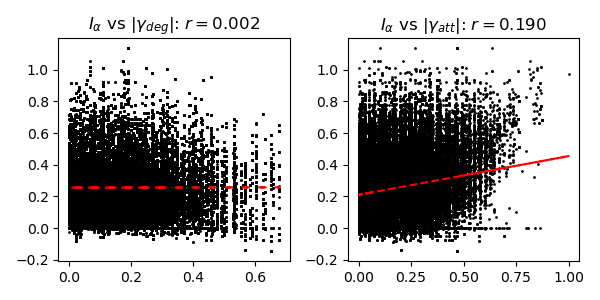}
	\caption{(Claim 3 of Section \ref{subsec: why Delta I}) The scatter plot reveals that the proposed measure $I_{\alpha}$ has a weak correlation with degree assortativity due to the effects of $\Ione$ and $\Itwo$ canceling out, while the correlation between $I_{\alpha}$ and $\atwo$ is significantly stronger than that between $\Itwo$ or $\Ione$ and $\atwo$. Thus, it can be concluded that $I_{\alpha}$ effectively captures the node attribute information of a network.}
	\label{fig: dI correlation}
\end{figure}

We demonstrate the correlation between the proposed \metric{} and $\aone$ and $\atwo$ in Figure \ref{fig: dI correlation}. Because both $\Ione$ and $\Itwo$ are positively correlated with $|\aone|$, their cancellation partially leads to $I_{\alpha}$ having a relative weaker association with $|\aone|$. More significantly, the correlation between $I_{\alpha}$ and $|\atwo|$ strengthens.

We further explain the efficacy of $I_{\alpha}$ by referring to the example in Figure \ref{fig: counter example}. Note that $\Ione$ (\ref{eq: degree MI expanded}) is basically the divergence between the joint degree distribution $e_{kk'}$ and the product of the marginal distributions $q_k \otimes q_{k'}$. Similarly, $\Itwo$ measures the divergence between the joint distribution $p(k, k', c, c')$ and $(q_k, c_k) \otimes (q_{k'}, c_{k'})$. The concentrated entries of $p(k, k', c, c')$, as demonstrated by the dark red cells of the $\hat{\jdam}$, cannot be adequately approximated by the product of the 1-dimension marginal distributions, thus increasing $I_{\alpha}$ significantly.

\vspace{0.2in}

The proposed \metric{} stands out in its ability to distinguish structural network inequalities that assortativity and homophily alone are unable to identify, as demonstrated through the three claims in Section \ref{subsec: why Delta I}. By improving the information provided by the node attributes and capturing the cancellation and synergy effects of $\Ione$ and $\Itwo$, it offers a comprehensive approach to analyzing network structures.

Building on this foundation, our analysis reveals that the Rényi mutual information of order 1.3 provides the highest correlation coefficient with $|\atwo|$, making it the most expressive of the node attribute information. This finding opens up exciting new avenues for more efficient and effective ways of describing complex information in networks.

In the next Section, we demonstrate that the proposed measure is also highly optimized and easy to use. These make it a valuable tool for researchers and practitioners in various fields who seek to understand and analyze network structures with greater precision and depth.

\section{Stochastic Optimization for Reducing the Network Inequality}\label{sec:optim}
As discussed in Section \ref{sec:information}, \metric{} measures the degree of inequality in an attributed network. 
In this section, we aim to reduce the network inequality by optimizing \metrictwo{}. 
Our approach involves using a parameterized distribution to sample the edges to be added to the network. By updating the parameterized sampling distribution through stochastic optimization, we aim to maximize the expected value of the mutual information measure. This approach is superior to deterministic edge addition or removal as it guarantees a local optimum, and it is effective in reducing inequalities such as the glass ceiling effect. It achieves this objective through random link recommendation, which preserves privacy \cite{cummings2019compatibility}.

\begin{figure}
	\centering
	\includegraphics[width=0.5\textwidth]{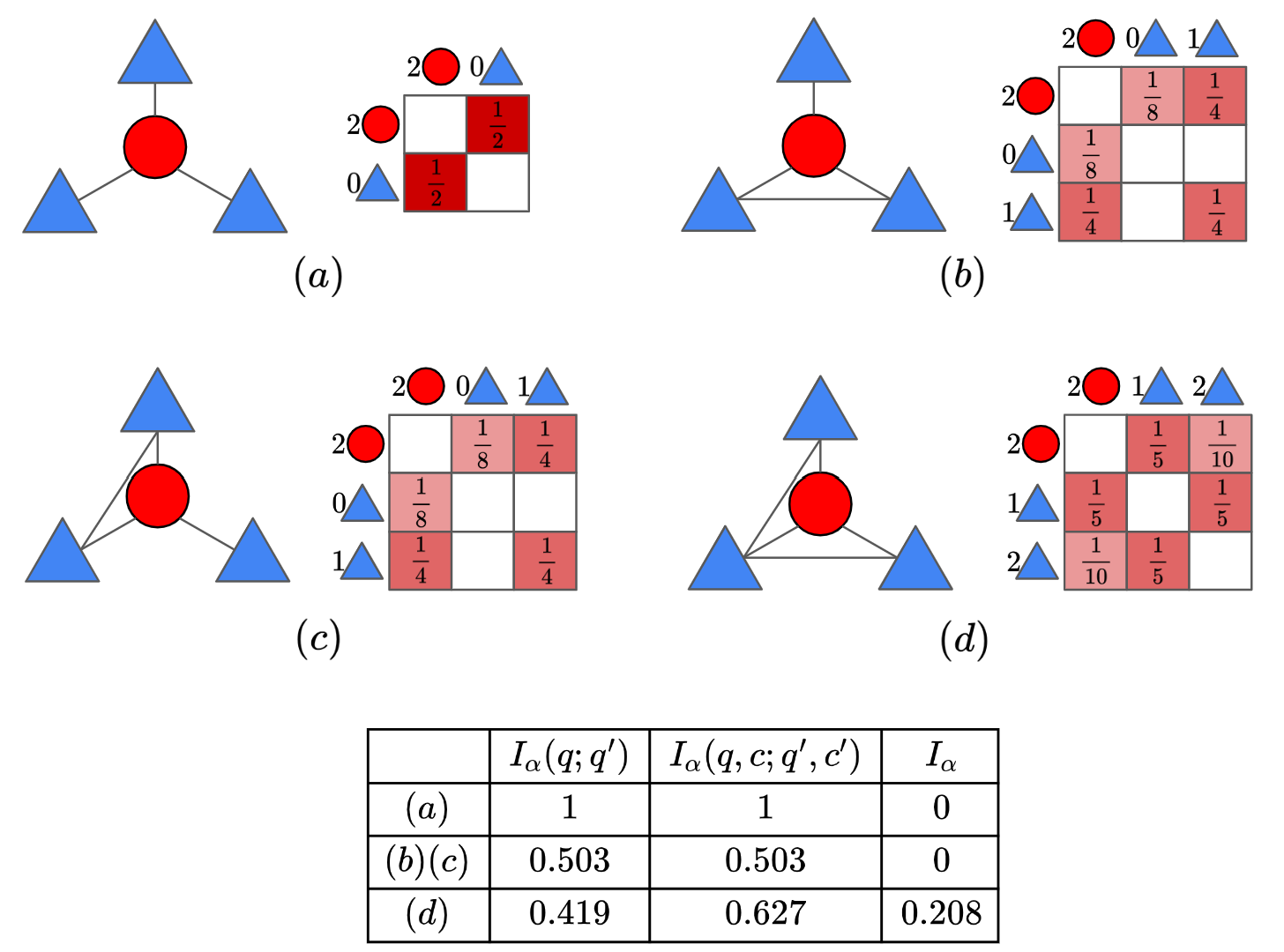}
	\caption{An example showing that the discrete optimization problem (\ref{eq:optim}) is not submodular. Therefore the performance of greedy algorithm cannot be guaranteed. We relax the problem and optimize over the space of edge addition using a stochastic optimization algorithm (Algorithm \ref{alg:SPSA}).
 } 
	\label{fig: submodular counter example}
\end{figure}

\subsection{Stochastic Optimization Problem Formulation}
In order to optimize the proposed mutual information measure of a network, we consider adding an edge between nodes in degree-attribute groups $(k, c)$ and $(k', c')$. By doing so, we increase the entry $[(k+1, c), (k'+1, c')]$ in \jdam{} by one while reducing the entry $[(k, c), (k', c')]$ by one\footnote{We also change the corresponding entries on the other side with respect to the diagonal of \jdam{} to preserve its symmetry.}. The set of possible degree-attribute groups before and after the edge addition is $\joint = {0, \cdots, \kmax} \times {+1, -1}$, which covers $|\joint| = 2(\kmax + 1)$ combinations. The edge space is $\jointedge = \joint \times \joint$, where $|\jointedge| = 4(\kmax + 1)^2$.

To formulate the optimization problem, we define the objective function as the change in \metrictwo{} resulting from adding an edge to the network, i.e., $\obj(x) = I_{\alpha}(E\cup \{ x\} ) - I_{\alpha}(E)$, where $x \in \jointedge$ and $E \cup \{x\}$ represents the edge set after the edge addition. However, we demonstrate with a counterexample (Figure \ref{fig: submodular counter example}) that the discrete optimization problem
\begin{equation}\label{eq:optim}
\begin{split}
    x^* \in \argmin\limits_{x\in \jointedge } \obj(x).
\end{split}
\end{equation}
is not submodular. Specifically, we find that $I_{\alpha}(E_a \cup \{x\}) - I_{\alpha}(E_a) = 0 < I_{\alpha}(E_c \cup \{x\}) - I_{\alpha}(E_c) = 0.208$, even though $E_a \subset E_c$. 

To overcome this challenge, we relax the problem into a stochastic optimization over probability mass functions, allowing us to find locally optimal solutions in a randomized manner. Specifically, we use a parameterized distribution to select which edges to add to the network and update the distribution via stochastic optimization to maximize the expected value of \metrictwo{}. This approach efficiently and effectively resolves the edge addition problem while reducing inequalities such as the glass ceiling effect.

Consider a family of probability mass functions ${f(x; \theta)}$ on $\jointedge$,  parameterized by the conditional logit model with separate fixed effect $\theta_i$ for each edge $i$:
\begin{equation}\label{eq:softmax}
    f_i(\jointedge; \theta) = \frac{\exp(\theta_i)}{\sum_{i=1}^{|\jointedge|} \exp(\theta_j)}
\end{equation}
We then formulate the optimization problem as
\begin{equation}\label{eq:new optimization}
\begin{split}
    \theta^* &= \argmin\limits_{\theta \in \mathbb{R}^{|\jointedge|}} C(\theta) \\
    &= \argmin\limits_{\theta \in \mathbb{R}^{|\jointedge|}} \mathbb{E}_{x\sim f(\jointedge; \theta)} \{Q(x)\}
\end{split}
\end{equation}
Here, we optimize the probability distribution of the edges in (\ref{eq:new optimization}), where the objective function is the expected increase in \metrictwo{} with respect to the probability distribution. 

In the following subsection, we explain how we use the Simultaneous Perturbation Stochastic Approximation (SPSA) algorithm \cite{spall1998implementation} to solve this stochastic optimization problem effectively. 

\begin{algorithm}
\caption{\textbf{SPSA based algorithm to estimate $\theta^*$}}\label{alg:SPSA}
\textbf{Input}: 
Initial parameterization $\theta^{(0)}$; \jdam{} of the current network.

\textbf{Output}: Estimate of the (locally) optimal solution $\theta^*$ of the conditional logit model for edge addition (\ref{eq:new optimization}). 

\begin{algorithmic}[1]
\For{$k=0, 1, \cdots$} %
    \State \multiline{Simulate the $|\jointedge|$-dimensional vector $d_k$ with random elements}
    \begin{equation}
        d_k(i) = \begin{cases}
            +1 & \; \textrm{with probability 0.5} \\
            -1 & \; \textrm{with probability 0.5}
        \end{cases}
    \end{equation}
    \State \multiline{Sample an edge $x$ from $\jointedge$ using $f(\jointedge; \theta^{(k)} + \Delta d_k)$ (\ref{eq:softmax}), where $\Delta > 0$.}
    \State \multiline{Compute the change in the \metrictwo{}, $C(\theta^{(k)} + \Delta d_k)$.}
    \State \multiline{Sample an edge $x$ from $\jointedge$ using $f(\jointedge; \theta^{(k)} - \Delta d_k)$.}
    \State \multiline{Compute $C(\theta^{(k)} - \Delta d_k)$.}
    \State Obtain the gradient estimate using (\ref{eq:gradient estimate}).
    \State Update $\theta^{(k)}$ via stochastic gradient descent
    \begin{equation}
        \theta^{(k+1)} = \theta^{(k)} + \epsilon \hat{\nabla}C(\theta^{(k)})
    \end{equation}
\EndFor
\end{algorithmic}
\end{algorithm}

\subsection{Stochastic Optimization Algorithm}
We use the Simultaneous Perturbation Stochastic Approximation (SPSA) algorithm \cite{spall1998implementation, krishnamurthy2016partially} to estimate the gradient of the new objective function with respect to each component of $\theta$. One of the key advantages of SPSA is that it requires only two simulations of the objective function, regardless of the dimension $|\jointedge|$ of the optimization problem. 

Algorithm \ref{alg:SPSA} outlines the SPSA implementation. In the $k$-th iteration,  all elements of $\theta$ undergo a random perturbation. A vector $d_k$ made up of $\{+1, -1\}$ with a Bernoulli$(0.5)$ distribution is scaled by a factor\footnote{$\Delta$ affects the bias-variance trade-off as the bias in the derivative estimate (\ref{eq:gradient estimate}) is proportional to $\Delta^2$ whereas the variance is proportional to $1/\Delta^2$ \cite{krishnamurthy2016partially}.} $\Delta>0$ and added and subtracted from $\theta$. Then, by sampling the edge addition using the distribution represented by the two perturbed $\theta$, the objective function $C(\theta)$ (\ref{eq:new optimization}) is evaluated twice. 
The derivative of $C(\theta)$ with respect to $\theta$ is approximated by 
\begin{equation}\label{eq:gradient estimate}
    \hat{\nabla}C(\theta^{(k)}) = \frac{C(\theta^{(k)} + \Delta d_k) - C(\theta^{(k)} - \Delta d_k)}{2\Delta} d_k 
\end{equation}
Finally, the stochastic gradient algorithm can be used to update $\theta$. As the SPSA algorithm employs stochastic gradient descent (\ref{eq:gradient estimate}), and the objective function may not be convex, the algorithm can only converge to a local stationary point. However, in practice, local optima can still provide useful solutions. Specifically, SPSA converges to a sampling distribution that is a local maximum of the objective function. 

In Section \ref{subsec:numerical optim}, we demonstrate the practical effectiveness of our approach through numerical results.

\section{Mutual Information Measure For Quantifying Glass Ceiling Effect in Citation Networks} \label{sec: numerical} %
In this section, we demonstrate the ability of \metrictwo{} to quantify the glass ceiling effect in citation networks, using both analytical models and real-world datasets. 
We use Algorithm \ref{alg:SPSA} from Section \ref{sec:optim} to optimize the distribution over edge addition, with the goal of reducing glass ceiling effect as measured by \metrictwo{}. Our results demonstrate that the optimized conditional logit model outperforms uniformly random edge addition, with a significant improvement in the reduction of network inequality. These findings highlight the importance of considering the distribution of edge additions in reducing the glass ceiling effect in citation networks. The code and datasets used in the experiments are publicly available at \url{https://tinyurl.com/mutual-information}.

\subsection{Directed Mixed Preferential Attachment Model}
We presented the Directed Mixed Preferential Attachment (DMPA) model \cite{nettasinghe2021emergence} in this subsection. We also employed a condensed version of it to validate the \metrictwo{} proposed in Section \ref{sec:information}.  

The DMPA model examines glass ceiling effect by modeling connection patterns in a growing citation network based on preferential attachment, minority group and homophily, as well as the rate at which new authors join. The two node types, $m$ and $f$ where $f$ represents the minority group,  correspond to the binary labels $c\in \{+1, -1\}$ defined in Section \ref{subsec: definition of coauthorship network}. The network begins with two connected nodes with different labels. At each time step, one of the three edge addition events occurs:

\vspace{0.1in}
\noindent (1) With probability $p$, a new node appears and an existing node cites it. The new node is assigned type $f$ with probability $p(f)\leq \frac{1}{2}$ and $m$ with probability $1-p(f)$. The potential citing node is chosen from the existing nodes with probability proportional to their in-degrees plus a constant $\delta$. A new citation edge is then created %
based on the probability matrix: 
\begin{equation}
\label{eq: Patt}
P_{\textrm{att}} =
    \begin{bmatrix}
        \homoprob & 1 - \homoprob \\
        1 - \homoprob & \homoprob
    \end{bmatrix}
\end{equation}
Specifically, an edge is added with probability $\homoprob \in (0, 1)$ if both nodes have the same type. If not, an edge is added with probability $1-\homoprob$. A value of $\homoprob > 0.5$ indicates homophily whereas $\homoprob < 0.5$ indicates heterophily. We simplify the model by assuming nodes with different types share the $\homoprob$, making $P_{\textrm{att}}$ symmetric.

\vspace{0.1in}
\noindent (2) With probability $q$, a new node appears and cites an existing node. The new node's type is assigned in the same manner as in case (1). The potential cited node is chosen from the existing nodes with probability proportional to their out-degrees plus a constant $\delta$. Then, a new citation edge is created based on $P_{\textrm{att}}$ (\ref{eq: Patt}).

\vspace{0.1in}
\noindent (3) With probability $1-p-q$, a new edge is created between two existing nodes. Both the citing and cited nodes are chosen independently based on their in- and out-degree, and they are connected based on $P_{\textrm{att}}$ (\ref{eq: Patt}).

\begin{figure}
	\centering
	\includegraphics[width=0.4\textwidth]{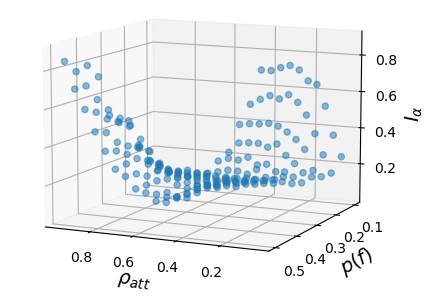}
	\caption{The \metrictwo{} $I_{\alpha}$ of DMPA networks that were simulated from various combinations of $p(f)$ and $\homoprob$.  $p(f)$ - the proportion of the female nodes; $\homoprob$ - the probability that two nodes with the same gender connect with each other. Both homophily and heterophily have a tendency to raise the \metrictwo{}, according to the scatter plot's U-shape.}
	\label{fig: DMPA}
\end{figure}

\begin{figure}
	\centering
	\includegraphics[width=0.45\textwidth]{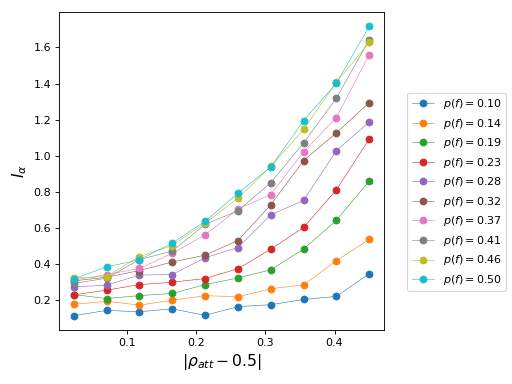}
	\caption{The relationship between $I_{\alpha}$ and $|\homoprob -0.5|$ of networks simulated from the DMPA model. Different colors represent different proportions of female nodes $p(f)$. The positive correlation supports our claim that higher homophily or heterophily causes more \metrictwo{} to be placed on the node attributes.}
	\label{fig: DI vs rho}
\end{figure}

\vspace{0.1in}
In this condensed DMPA model above, $p(f)$ controls the imbalance between $m$ and $f$, $\homoprob$ determines whether the nodes are homophilic or heterophilic and is closely related to the attribute assortativity $\atwo$ (\ref{eq: atwo}). When $\homoprob=0.5$, a node is indifferent between choosing nodes with the same type or a different type. In addition, $p, q$ control the relative frequency with which new nodes join the network. Last but not least, bigger $\delta > 0$ corresponds to weaker preference to high-degree nodes under the preferential attachment.

In the simulation\footnote{We used the code provided by authors of \cite{nettasinghe2021emergence}: \url{https://github.com/ninoch/DMPA}.}, we set $p(f) \in (0.1, 0.5)$ so that type $f$ corresponds to the minority. We vary $\homoprob$ from 0.05 to 0.95 which covers a spectrum of homophily and heterophily. We set $p=q=0.15$ and $\delta=10$. For each combination of the parameters $\{p(f), \homoprob\}$, we generate a network with 10000 edges, which have approximately 3000 nodes.   

Figure \ref{fig: DMPA} displays $I_{\alpha}$s of networks simulated using various combination of $\{p(f), \homoprob\}$. The U-shape of the scatter plot shows that both homophily and heterophily tend to raise the \metrictwo{}. 
We further plot $I_{\alpha}$ against $|\homoprob - 0.5|$ under different $p(f)$ in Figure \ref{fig: DI vs rho}, where $|\homoprob - 0.5|$ measures how homophilic or heterophilic the network is. The positive correlation between $I_{\alpha}$ and $|\homoprob - 0.5|$ supports our argument that higher homophily or heterophily results in more attention paid to the node attributes. %

\subsection{Empirical Analysis of Publication Data}
\label{subsec: empirical}
To confirm that \metrictwo{} is a useful measurement for describing the glass ceiling effect in real-world data, we use the publishing records from public datasets. 
We use the bibliographic data of publications in the field of Physics, Management, Psychology, and Political Science \cite{nettasinghe2021emergence}. %
The authors' genders are extracted using third-party APIs\footnote{\url{https://namsor.app/}; \url{https://www.gender-api.com/}.}.

\begin{figure}
	\centering
	\includegraphics[width=0.5\textwidth]{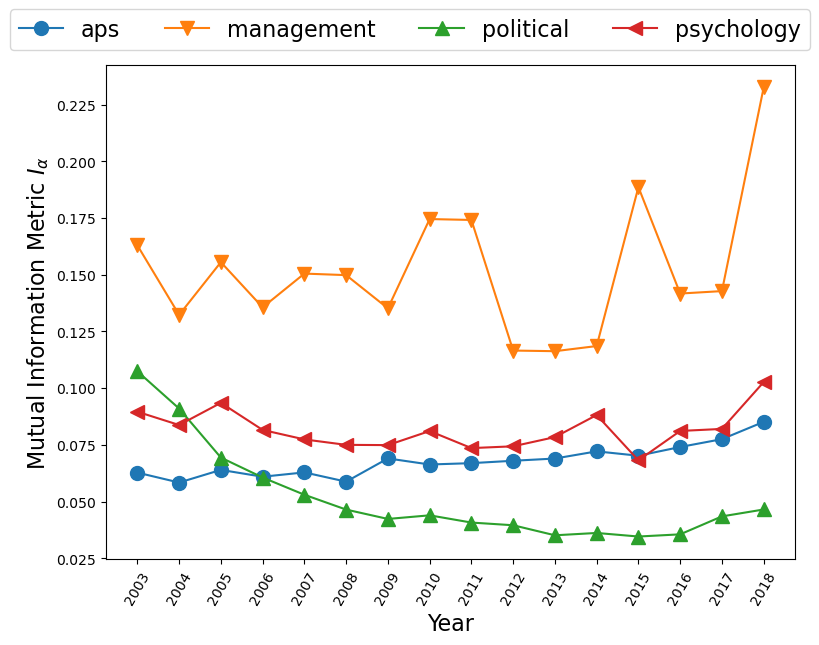}
	\caption{Temporal evolution of the \metrictwo{} $I_{\alpha}$ for four different fields, including Physics, Management, Psychology, and Political Science. The plot shows the changes in $I_{\alpha}$ over time, where decreasing values of $I_{\alpha}$ indicate a reduction in the glass ceiling effect. The results indicate that in Political Science, $I_{\alpha}$  steadily decreases over time, suggesting reduced gender bias in citation formation. In contrast, $I_{\alpha}$  shows a consistent growth over time for Physics and Psychology, indicating an increasing glass ceiling effect as gender becomes a more significant factor in citation formation. On the other hand, $I_{\alpha}$ remains relatively stable over time for Management, except for a sudden increase in 2018, possibly due to incomplete citation data.
 } 
	\label{fig: infogain citation}
\end{figure}

We construct citation networks as follows. We group the papers in each field based on the publication year to build the network per field per year. Because the dataset for the field of Physics comprises of papers published before 2019, we build citation network for all the 5 fields from 2003 to 2018. We view citation edges as undirected (See Section \ref{subsec: definition of coauthorship network} for definition of a citation network) so that the adjacency matrix and the $\jdam$ are symmetric.

Figure \ref{fig: infogain citation} presents the \metrictwo{} values for the four fields analyzed in this study. The results show that, in Political Science, \metrictwo{} gradually decreases over time, indicating that authors have paid less attention to gender as a factor in citation formation. However, for Physics and Psychology, \metrictwo{} exhibits a steady increase over the years, suggesting that the glass ceiling effect has grown in these fields as gender has become a more significant determinant of citation patterns. In contrast, the glass ceiling effect in Management appears to be generally consistent over the years, with the exception of a late jump in the year 2018, possibly due to incomplete citation data. These findings provide valuable insights into the evolution of gender-based disparities in citation practices across different academic fields.

\subsection{Optimization on Randomized Edge Addition}\label{subsec:numerical optim}
Previous numerical experiments on real-world citation networks show that \metrictwo{} $I_{\alpha}$ depicts the glass ceiling effect of a network over time. 
We implement Algorithm \ref{alg:SPSA} in Section \ref{sec:optim} and show that the $I_{\alpha}$  can be optimized using randomized edge addition. 

We use the citation network in physics from year 2012 as an example. As can be seen in Figure \ref{fig: infogain citation}, $I_{\alpha}$  is relatively high. 
We use a parameterized conditional logit model (\ref{eq:softmax}) to select edges to add to the network. $\theta_i, i=1, \cdots, |\jointedge|$ is initialized using a standard normal distribution. We run the algorithm for 10000 iterations. 

\begin{figure}
	\centering
	\includegraphics[width=0.5\textwidth]{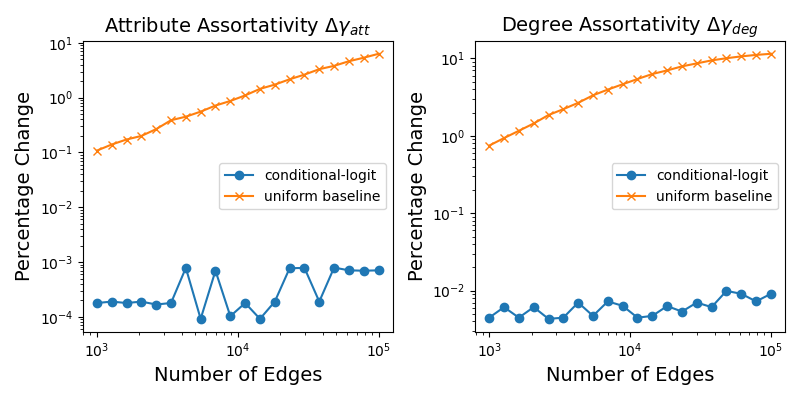}
	\caption{Effect of adding edges on attribute assortativity ($\atwo$) and degree assortativity ($\aone$) of the physics citation network from 2012. The x-axis represents the number of edges added to the network, and both axes are shown on a logarithmic scale. The plots show that the optimized conditional logit model results in a greater increase in both attribute and degree assortativity compared to the baseline method that uses a uniform distribution of edge addition.
 } 
	\label{fig: edge addition}
\end{figure}

We add various numbers of edges to the original network using the conditional logit model to evaluate the optimization result. The baseline method adds the same amount of edges but uses a uniform distribution. Figure \ref{fig: edge addition} shows that both the attribute assortativity $\atwo$ and the degree assortativity $\aone$ increase more in the baseline method compared with the optimized conditional logit model. 
As such, by recommending links in a randomized fashion, optimization on \metrictwo{} could reduce inequality in a network while preserving the privacy of the network participants.

\section{Conclusions and Extensions}
\label{sec:conclusion}
\noindent
{\bf Conclusions: }
This paper quantifies the glass ceiling effect in networks using the mutual information measure (based on Shannon and more generally, Rényi entropy) between the conditional probability distributions of node attributes given the degree of adjacent nodes. Compared to existing measures, such as degree assortativity and homophily, \metrictwo{} accounts for both demographic information and node degrees, making it a more comprehensive measure of node attribute information in networks. We demonstrate the effectiveness of \metrictwo{} through examples and simulations. %
We then apply \metrictwo{} to citation networks and demonstrate its ability to identify structural inequality and track the evolution of the glass ceiling effect across different research fields over time. Our results suggest strategies to support minority scientists and demonstrate the potential of \metrictwo{} to contribute to a better understanding of glass ceiling effects in networks.

\vspace{0.2in}
\noindent
{\bf Limitations and Extensions: }
While the proposed \metrictwo{} provides a comprehensive measure of node attribute information in networks, there are some limitations that suggest avenues for future research. Firstly, when the range of degree distributions differs, the proposed method may not be able to compare various networks effectively. To address this issue, we can apply degree cutoffs to resolve the problem of a broad degree distribution.
Secondly, \metrictwo{} quantifies the network as a whole, while ignoring differences in different regions of the network. A promising direction for future research is to use local assortativity\cite{piraveenan2009local} to improve the measure's expressiveness in network subgraphs.
Another potential avenue for future research is to aggregate the information from coauthorships and citations using multiplex network models where nodes are linked in multiple interacting layers\cite{menichetti2014weighted}. In particular, we can generalize the joint degree and attribute matrix to account for the correlation between weights and topology of the multiplex networks.

\ifCLASSOPTIONcaptionsoff
  \newpage
\fi

\bstctlcite{IEEEexample:BSTcontrol}
\bibliographystyle{IEEEtran}

\end{document}